# Enabling Self-Powered Autonomous Wireless Sensors with New-Generation I$^2$C-RFID Chips

Danilo De Donno, Luca Catarinucci, and Luciano Tarricone

University of Salento, Innovation Engineering Department
Via per Monteroni, 73100, Lecce – Italy
mail:{danilo.dedonno; luca.catarinucci; luciano.tarricone}@unisalento.it

*Abstract* — A self-powered autonomous RFID device with sensing and computing capabilities is presented in this paper. Powered by an RF energy-harvesting circuit enhanced by a DC-DC voltage booster in silicon-on-insulator (SOI) technology, the device relies on a microcontroller and a new generation I$^2$C-RFID chip to wirelessly deliver sensor data to standard RFID EPC Class-1 Generation-2 (Gen2) readers. When the RF power received from the interrogating reader is -14 dBm or higher, the device, fabricated on an FR4 substrate using low-cost discrete components, is able to produce 2.4-V DC voltage to power its circuitry. The experimental results demonstrate the effectiveness of the device to perform reliable sensor data transmissions up to 5 meters in fully-passive mode. To the best of our knowledge, this represents the longest read range ever reported for passive UHF RFID sensors compliant with the EPC Gen2 standard.

## I. Introduction

In recent years, the ultra-high-frequency (UHF) radio frequency identification (RFID) technology is expanding its range of applications by taking advantage from sensor-augmented RFID tags. As advocated in [1], the RFID sensor network (RSN) concept extends the standard RFID functionalities beyond the mere identification and tracking [2]-[4] to support ubiquitous computing and sensing. Not only is the research activity very intense in this subject field, but also a large variety of commercial products and solutions is rapidly emerging.

As far as the research efforts are concerned, the most attractive RFID device integrating sensing and computing capabilities is undoubtedly the Intel WISP (wireless identification and sensing platform) [5]. The WISP is a fully passive and programmable UHF RFID tag with sensors that implements the EPC Class-1 Generation-2 (Gen2 for short hereafter) protocol. A low-cost alternative to the WISP is the sensor-tag (S-Tag) [6]. Based on the multi-ID approach, the S-Tag can be connected to generic sensors and, when interrogated by a standard Gen2 reader, it is capable to transmit a proper combination of EPC codes that univocally encodes the sensor value. Different approaches are proposed in [7] and [8] where the electronic components of an UHF RFID tag, including antenna, microcontroller unit (MCU), and sensors, are integrated in organic substrates.

Among the commercial RFID Gen2 tags with sensors, the most interesting are the SL900A sensory tag by IDS Microchip [9] and the Easy2Log tag by CAEN RFID [10]. As for the former, read ranges up to 1.5 meters and 5.2 meters are reported respectively for the fully-passive and semi-passive (battery-assisted) modes of operation. The latter, instead, can operate solely with battery assistance exhibiting a maximum read range of 10 meters.

In this work, we rely on a novel approach, never reported before, to design and prototype an RFID-augmented module with sensing and computing capabilities. The developed device, fabricated using off-the-shelf low-cost discrete components, is self-powered by an RF energy-harvesting circuit enhanced by a DC-DC booster in silicon-on-insulator (SOI) technology. An on-board ultra-low-power MCU manages the sensor data sampling and the wireless communication by means of a new generation UHF I$^2$C-RFID chip whose EPC code is dynamically updated with actual sensor measurements. The prototyped device is able to autonomously produce rectified DC voltages of 2.4 V from received RF power levels as low as -14 dBm, thus enabling RFID-based sensor data transmission up to 5 meters.

## II. System architecture

The block diagram of the proposed RFID-based sensor module is depicted in Fig. 1 while the developed printed circuit board (PCB) prototype is shown in Fig. 2. A 50-Ω dipole-like antenna, an LC impedance-matching network, and an RF-DC rectifying circuit compose the RF energy-harvesting section. The rectified DC voltage is boosted by a DC-DC charge-pump IC which also stores and efficiently releases energy to power up the digital section. Here, an ultra-low power MCU samples embedded and/or external sensors and transfers the data to a new-generation UHF RFID chip featuring I$^2$C interface. In this work, as a simple proof of concept, the case of ambient temperature sensing is considered. Finally, the sensor data are delivered to the interrogating reader via the UHF RFID Gen2 communication protocol. It is worth remarking that the implemented RFID-based sensor module is completely passive. In fact, it harvests and converts into usable energy for its operation the same RF power transmitted by the reader for the chip interrogation.

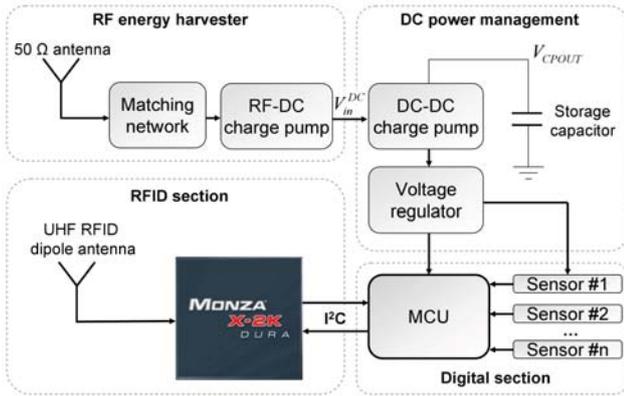

Fig. 1. Block diagram of the proposed RFID-based sensor node.

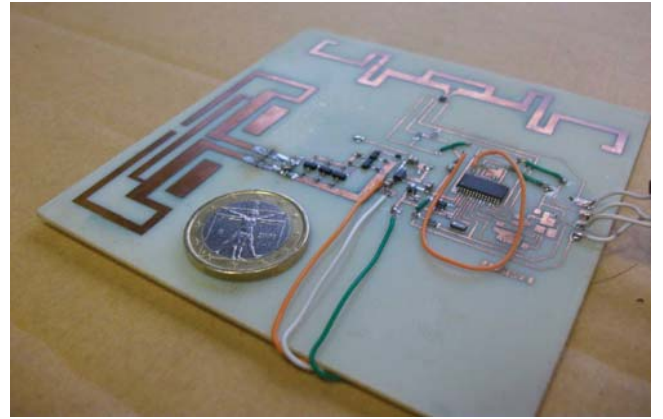

Fig. 2. Prototype photo. Overall dimensions: 9.5x9.5x0.16 cm$^3$.

The main parts making up the system architecture are thoroughly described in the following subsections.

*A. RF energy harvesting and DC power management*

The adopted RF-DC converter is a 5-stage Cockcroft-Walton voltage multiplier [11] that is matched to a 50-Ω dipole antenna by an LC matching network. Both rectifier and antenna were tuned at 866.5 MHz, i.e. the center frequency of the European UHF RFID band. The Skyworks SMS7630 [12] zero-bias Schottky diodes have been chosen as rectifying devices because of their high-detection sensitivity at UHF frequencies. The Seiko Instruments S-882Z charge pump IC [13] is then used to step-up the rectified voltage $V_{in}^{DC}$. Such a DC-DC converter adopts fully depleted SOI technology to enable ultra-low voltage operation. In fact, when $V_{in}^{DC}$ is 0.35 V or higher the oscillation circuit starts to operate and the stepped-up electric power is accumulated in a storage capacitor. When the capacitor reaches $V_{CPOUT}$ =2.4 V, the stored voltage is supplied to a voltage regulator to power up the digital section.

*B. Digital section*

An ultra-low power 16-bit MSP430 MCU [14] by Texas Instruments (TI) with integrated 10-bit analog-to-digital converter (ADC) is programmed to sample a TI LM940 [14] analog temperature sensor. The code running on the MCU also implements I$^2$C communication routines to write the sensor readings into the memory banks of the new Monza X-2K [15] I$^2$C-RFID Gen2 chip by Impinj.

When $V_{CPOUT}$ declines to approximately 1.85 V as a result of the storage capacitor discharge, the S-882Z IC automatically stops the discharge process and, consequently, the MCU code execution. The idle time between MCU operations, i.e. the duty cycle of the overall system, is determined by the amount of input power to the charge-pump IC to charge up the storage capacitor. In the proposed application, it is experimentally found that a 10 μF storage capacitor is sufficient for the temperature sensing and I$^2$C communication tasks.

*C. RFID section*

In the RFID section of the system, we designed a dipole-like antenna and tuned its complex input impedance to match that of an Impinj Monza X-2K RFID chip ($Z_{chip} = 20.83 - j181.39\ \Omega$). The Monza X-2K is an UHF RFID Gen2 IC with 2176 bits of non-volatile memory (NVM) and an I$^2$C interface. As an I$^2$C device, Monza X-2K operates as a standard I$^2$C EEPROM whose contents can also be accessed wirelessly via the Gen2 protocol. We programmed the MCU in the digital section to write the 10-bit temperature samples in the EPC memory banks of the Monza X-2K. In this way, since the EPC of an RFID tag, i.e. its ID number, is the easiest and most valuable information accessible to the users, the temperature data can be displayed and managed using standard off-the-shelf readers.

III. EXPERIMENTAL RESULTS

The designed RFID-based sensor node was fabricated using off-the-shelf low-cost discrete components on an FR4 substrate. Antennas for both the RF energy harvester and the Monza X-2K RFID IC were patterned directly on the PCB as shown in Fig. 2. In a first set of experiments, we conducted a sensitivity analysis in anechoic chamber. A software-defined radio (SDR) equipment [16] was connected to a circularly polarized antenna with gain $G_{tx}$ =5.5 dBi and placed at 1 meter of distance from the device. According to the measurement procedures described in [17]-[21], the transmit power was gradually increased in order to find the minimum value $P_{tx,ON}$ required to power up the MCU and retrieve correct temperature data. The measurements were repeated at different frequencies in the 840-900 MHz band. Then, the sensitivity was calculated by the following equation based on the free-space Friis' propagation model:

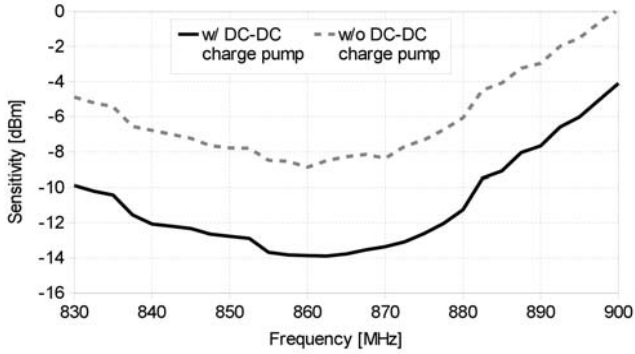

Fig. 3. Measured sensitivity of the proposed RFID-based sensor node with and without the DC-DC charge pump.

$$S = EIRP_{ON} G_{rx} \left(\frac{\lambda}{4\pi d}\right)^2 \eta_{plf} \quad (1)$$

where $EIRP_{ON} = P_{tx,ON} G_{tx}$ is the minimum equivalent isotropically radiated power (EIRP) required to energize the device, $G_{rx}$ =1.8 dBi is the gain of the 50-Ω energy-harvesting antenna, $\lambda$ is the wavelength, $\eta_{plf}$ =0.5 is the polarization loss factor due to the circularly polarized transmit antenna, and $d$ =1 meter is the distance between the interrogating antenna and the device under test. The achieved results are plotted in Fig. 3 where the gain yielded by the DC-DC charge pump over the sole rectifier is also shown. The proper tuning of the device is demonstrated by the better sensitivity values achieved around 866.5 MHz (-14 dBm and -9 dBm respectively for the circuit with and without the DC-DC charge pump).

In a second set of experiments, the performance of the RFID-sensor node was evaluated in real operating conditions. In Fig. 4, we plot the average number of temperature reads per minute when varying the distance from a commercial UHF RFID reader set with the maximum allowable transmit power for the European regulations, i.e. 3.2-W EIRP in the 865–868 MHz frequency band. The experiments were conducted in a standard office room with reader antenna and device under test placed along a straight line 1.5 meters above the floor. It is clear from Fig. 4 that, in the 0–1.5 meters range, the duty cycle introduced by the DC-DC charge pump causes a lower read rate compared to the case where the DC-DC booster is bypassed. However, as expected from the sensitivity analysis, the DC-DC charge pump allows to significantly extend the operating range up to 4.8 meters, i.e. approximately three times greater than the range achieved without DC-DC boosting.

Finally, Fig. 5 demonstrates the temperature measurement capabilities of the developed RFID-based sensor node. The experimental results show a nearly perfect agreement with the temperature values provided by a commercial DigiTemp probe [22] used as reference. The maximum error recorded was 1.3 °C.

IV. CONCLUSION

This paper has presented the design and experimental validation of a self-powered wireless sensor module compliant with the UHF RFID Gen2 standard. The device uses an RF-DC rectifier boosted by a DC-DC charge pump in SOI technology to harvest energy from the interrogating reader and power up a microcontroller, sensors (a thermometer in the presented case), and a new generation I$^2$C-RFID chip whose EPC code is dynamically updated with actual sensor measurements. Our preliminary prototype starts to operate when the received RF power level is as low as -14 dBm, thus enabling RFID-based sensor data transmission up to approximately 5 meters, which is, to the best of our knowledge, the longest ever reported distance for similar devices. In the progress of this work, we plan to further enhance the energy-harvesting capabilities by means of more sophisticated RF-DC converters and additional free-energy sources, e.g. solar, thermal, etc. Data-logging functionalities will be also included.

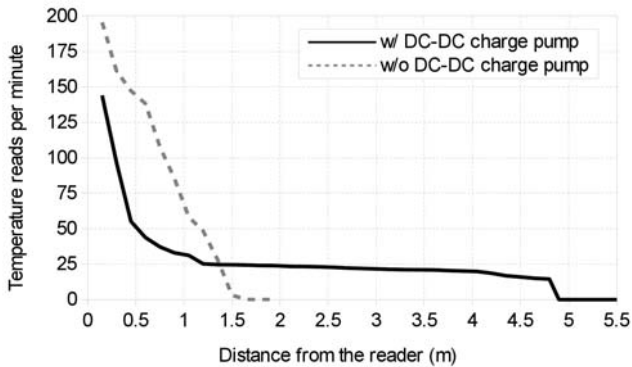

Fig. 4. Temperature data transmission rate vs. distance from the interrogating UHF RFID reader.

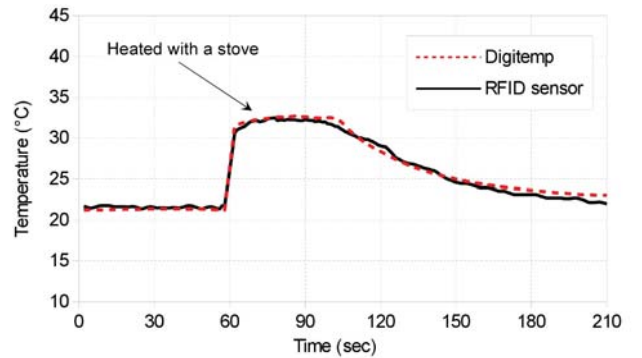

Fig. 5. Temperature measurements: comparison between a wired Digitemp probe and our RFID-based sensor node placed at 1 meter of distance from the interrogating reader.